\documentclass[twocolumn,showpacs,preprintnumbers,aps,prl]{revtex4}
\usepackage{graphicx}
\usepackage{dcolumn}
\usepackage{bm}
\begin{document}
\draft \preprint{TbMn$_2$O$_5$}
\title{Non-resonant and Resonant X-ray Scattering Studies on Multiferroic TbMn$_2$O$_5$\\}
\author{J. Koo$^1$}
\author{C. Song$^1$}
\author{S. Ji$^1$}
\author{J.-S. Lee$^1$}
\author{J. Park$^1$}
\author{T.-H. Jang$^1$}
\author{C.-H. Yang$^{1}$}
\author{J.-H. Park$^{1,2}$}
\author{Y. H. Jeong$^1$}
\author{K.-B. Lee$^{1,2}$}
\email{kibong@postech.ac.kr}
\author{T.Y. Koo$^2$}
\author{Y.J. Park$^2$}
\author{J.-Y. Kim$^2$}
\author{D. Wermeille$^{3}$}
\altaffiliation[Present address: ]{European Synchrotron Radiation
Facility, BP 220, F-38043 Grenoble Cedex 9, France}
\author{A.I. Goldman$^3$}
\author{G. Srajer$^4$}
\author{S. Park$^5$}
\author{S.-W. Cheong$^{5,6}$}
\affiliation{$^1$eSSC and Department of Physics, POSTECH, Pohang 790-784, Korea\\
$^2$Pohang Accelerator Laboratory, Pohang University of Science and
Technology, Pohang 790-784, Korea\\$^3$Ames Laboratory, Department
of Physics and Astronomy, Iowa State University, Ames, IA 50011,
USA\\$^4$Advanced Photon Source, Argonne National Laboratory, Argonne, IL 60439, USA\\
$^5$Rutgers Center for Emergent Materials and Department of Physics
and Astronomy, Rutgers University, Piscataway, NJ 08854,
USA\\
$^6$Laboratory of Pohang Emergent Materials and
Department of Physics, POSTECH, Pohang 790-784, Korea }

\date{\today}

\begin{abstract}
Comprehensive x-ray scattering studies, including resonant
scattering at Mn \textit{L}-edge, Tb \textit{L}- and
\textit{M}-edges, were performed on single crystals of
TbMn$_2$O$_5$. X-ray intensities were observed at a forbidden Bragg
position in the ferroelectric  phases, in addition to  the lattice
  and the magnetic modulation peaks.
Temperature dependences of their intensities and
the relation between the modulation wave vectors provide direct
evidences of exchange striction induced ferroelectricity.
Resonant
x-ray scattering results demonstrate the presence of multiple magnetic
orders by exhibiting their different temperature dependences.
The commensurate-to-incommensurate phase transition around 24 K  is
attributed to discommensuration through phase slipping of the
magnetic orders in spin frustrated geometries.
We proposed that the low temperature incommensurate
phase consists of the commensurate magnetic domains
separated by anti-phase domain walls which reduce spontaneous polarizations
abruptly at the transition.

\end{abstract}
\maketitle

In recent years, much attention has been paid to multiferroic
materials, in which magnetic and ferroelectric orders coexist and
are cross-correlated \cite {1,2,3,4,5,6,7,8,9,10}, due to
theoretical interests and potential application to  magnetoelectric
(ME) devices. Manipulation of electric polarizations by external
magnetic fields has been demonstrated in some of these materials
\cite{4,5}. Orthorhombic TbMn$_2$O$_5$, one of the multiferroic
materials, displays a rich phase diagram. Upon cooling through
\textit{T}$_N$ $\sim$ 41 K, TbMn$_2$O$_5$ becomes antiferromagnetic
with an incommensurate magnetic (ICM) order which transits to a
commensurate magnetic (CM) phase with spontaneous electric
polarization at \textit{T}$_{c1}$ $\sim$ 36 K, and reenters a low
temperature incommensurate magnetic (LT-ICM) phase at
\textit{T}$_{c2}$ $\sim$ 24 K. Anomalies of ferroelectricity and
dielectric properties were observed concurrently with these magnetic
phase transitions \cite{4,9}. Especially, the reentrant LT-ICM phase
is a phenomenon peculiar to \textit{R}Mn$_2$O$_5$ multiferroics
while commensurate phases are more common as the low temperature
ground states. Since the CM to LT-ICM phase transition is also
accompanied with an abrupt loss of spontaneous polarizations, it is
critical to elucidate the natures of the incommensurability of the
material, including the mechanism of the CM to LT-ICM phase
transition.

The origin of the complex phases of the material is attributed to
the coupling between magnetic moments of Mn ions and lattice
\cite{8,9}. It is suggested that, when a magnetic order is modulated
with a wave vector \textit{\textbf{q}$_m$}, the exchange striction
affects
 inter-atomic bondings resulting in a
periodic lattice modulation with a wave vector
\textit{\textbf{q}$_c$} = 2\textit{\textbf{q}$_m$} \cite{5,6,7,8,9}.
Recently, Chapon \textit{et al.} proposed for \textit{R}Mn$_2$O$_5$
systems that ferroelectricity results from the exchange striction of
acentric spin density waves for the CM phases \cite{9}. Indeed,
Kimura \textit{et al.} insisted that CM modulations are
indispensable to the ferroelectricity in the LT-ICM phase, from
their  neutron scattering results on HoMn$_2$O$_5$ under high
magnetic fields \cite{11}. However, lattice distortions derived from
ICM spin structures turned out to describe well the spontaneous
polarizations of YMn$_2$O$_5$ even in the ICM phase \cite{12},
implying that commensurability is not a necessary condition for the
ferroelectricity. In order to understand the intriguing
magnetoelectricity well, detailed information on  the lattice and
spin structure changes is necessary. However, only  limited
crystallographic data are available and even any direct evidence on
the symmetry lowering has not been reported yet
\cite{9,10,11,12,13,14}.

In this letter, we present synchrotron x-ray scattering results
on single crystals of TbMn$_2$O$_5$.
Since x-ray scattering is sensitive to both lattice
 and magnetic modulations, x-ray scattering with intense
undulator x-rays allowed simultaneous measurements for
\textit{\textbf{q}$_m$} and \textit{\textbf{q}$_c$}. Non-resonant
x-ray scattering  results show the relationship of
\textit{\textbf{q}$_c$} = 2\textit{\textbf{q}$_m$}, confirming
lattice modulations are generated by the magnetic orders. A (3 0 0)
forbidden Bragg peak, which is a direct evidence of the symmetry
lowering to a non-centrosymmetry space group, was observed in the
ferroelectric (FE) phases. Furthermore, the temperature dependence
of the peak intensity, \textit{I}$_{(300)}$, was found to coincide
with those of the lattice modulation peak intensities,
\textit{I}$_c$, and the spontaneous polarization
square,$\emph{P}^2$, in the CM phase. This indicates the
ferroelectricity is generated by the lattice modulations. In the
LT-ICM phase, temperature dependences of \textit{I}$_c$ cannot be
described by a single order parameter, implying the presence of
different magnetic orders. Resonant x-ray magnetic scattering
results at Mn \textit{L}-, Tb \textit{L$_3$}- and
\textit{M$_5$}-edges show that each magnetic order has its own
temperature dependence. It is proposed that  CM to LT-ICM phase
transition is induced by discommensuration through phase slipping
due to competing magnetic orders under the frustrated geometry.
Moreover, the CM modulations with anti-phase domain walls are
consistent with the temperature dependences of
\textit{\textbf{q}$_m$} and \textit{I}$_{(300)}$ in the LT-ICM
phase, and explain well the abrupt loss of $\emph{P}$ at the
transition.

Single crystals of TbMn$_2$O$_5$ were grown by a flux method
\cite{4}. The specimen used for the hard x-ray scattering
measurements has a plate-like shape with (1 1 0) as a surface normal
direction. Its mosaicity was measured to be about 0.01$^\circ$ at (3
3 0) Bragg reflection. For soft x-ray scattering, a different sample
was cut and polished to have (2 0 1) as a surface normal direction.
Soft x-ray scattering measurements were performed at 2A
beamline in the Pohang Light Source (PLS).
Details of the soft x-ray scattering chamber were described elsewhere
\cite{15}.
X-ray diffraction
experiments were conducted at the 3C2 bending magnet beamline in the PLS
and at the 6-ID undulator beamline in the Midwest Universities
Collaborative Access Team (MUCAT) Sector in the Advanced Photon
Source. For non-resonant x-ray scattering experiments, 6.45 keV was
selected as an incident x-ray energy below Mn \textit{K}-edge ($\sim
$ 6.55 keV). All the incident x-rays were $\sigma$-polarized and
PG(006)  was used to have a $\sigma$-to-$\pi$ channel
at Tb \textit{L$_3$}-edge.
\begin{figure}[t]
\begin{center}
\includegraphics[width=8.5cm]{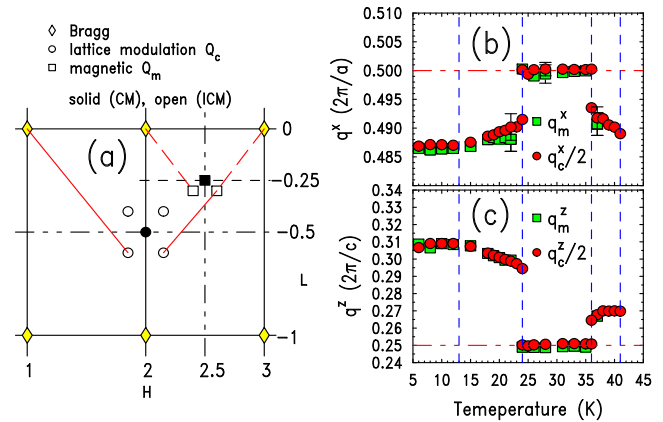}
\caption{(Color online) Positions of the measured magnetic
satellites (square) and lattice modulation peaks (circle) in the
($h$ 5 $l$) reciprocal lattice plane are shown in (a). The
temperature dependences of $q_m^x$ (square) and $q_c^x$ (circle),
and those of $q_m^z$ (square) and $q_c^z$ (circle) are shown in (b)
and (c), respectively. For direct comparisons with those of
\textit{\textbf{q}$_m$}, the components of \textit{\textbf{q}$_c$}
are divided by two. Vertical broken lines indicate \textit{T}$_{N}$
$\sim $ 41 K, \textit{T}$_{c1}$ $\sim$ 36 K, \textit{T}$_{c2}$ $\sim
$ 24 K and \textit{T}$_{c3}$ $\sim$ 13 K, respectively.}
\end{center}
\end{figure}

Nonresonant x-ray scattering measurements were performed to
investigate the temperature dependence of \textit{\textbf{q}$_m$}
and \textit{\textbf{q}$_c$} simultaneously. The measured lattice
modulation peak position of (2 5 -0.5) for the CM phase and those of
its 4 split peaks for the ICM phases are presented as solid and open
circles, respectively, in Fig. 1 (a). For magnetic satellites, (2.5
5 -0.25) peak and its 2 split ones were measured for the CM and ICM
phases. Their positions are presented as solid and open squares,
respectively. The magnetic and lattice modulation satellites for ICM
phases are linked with broken and solid lines to their corresponding
main Bragg peaks. Temperature dependences of \textit{\textbf{q}$_m$}
and \textit{\textbf{q}$_c$} are shown in Fig. 1 (b) and (c). From
the results, it is obvious that relation, \textit{\textbf{q}$_c$} =
2\textit{\textbf{q}$_m$}, holds within experimental errors in the
whole temperature range below \textit{T}$_N$. It is consistent with
the magnetic order induced lattice modulations. The temperature
dependence of \textit{\textbf{q}$_m$} shown here is qualitatively
similar to the neutron scattering results by others \cite{16}. Below
\textit{T}$_N$,  ICM magnetic peaks  develop, and
\textit{\textbf{q}$_m$} locks into a CM ordering at ($\frac{1}{2}$ 0
$\frac{1}{4}$) via a first order transition at \textit{T}$_{c1}$. On
further cooling the sample below \textit{T}$_{c2}$,
 the CM to  LT-ICM phase transition takes place.
With further decreasing temperature, \textit{\textbf{q}$_m$} of the
LT-ICM modulations evolves and is eventually pinned around (0.486 0
0.308) which can be approximated to a CM value of ($\frac{17}{35}$ 0
$\frac{4}{13}$) at \textit{T}$_{c3}$ $\sim$ 13 K. Such a long-period
CM modulation can be interpreted as the CM modulations
(\textit{\textbf{q}$_m$} = ($\frac{1}{2}$ 0 $\frac{1}{4}$)) with
domain walls, as is the case for ErNi$_2$B$_2$C \cite{17}.

\begin{figure}[b]
\begin{center}
\includegraphics[width=8.5cm]{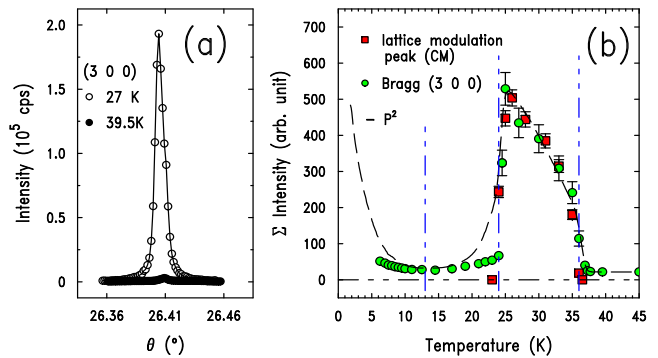}
\caption{(Color online) (a) Rocking curves of a (3 0 0) forbidden
Bragg peak measured below (open) and above \textit{T}$_{c1}$
(solid). (b) Temperature dependences of the integrated intensities
of a (3 0 0) Bragg peak (circle), CM lattice modulation peak
(square) and squared spontaneous polarization (broken line) taken
from Ref. 4. All the data are properly scaled. }
\end{center}
\end{figure}

As shown in Fig. 2 (a), measurable x-ray intensities were observed,
in the ferroelectric phase, at (3 0 0) Bragg position which is
forbidden under a space group of the room temperature paraelectric
phase, \textsl{Pbam}. Residual intensities above \textit{T}$_{c1}$
are due to higher harmonic contaminations. Values for
full-width-at-half-maximum (FWHM) of the peak are about
0.01$^\circ$, close to those of (4 0 0) main Bragg peak in the
LT-ICM phase. The results explicitly evidenced that inversion
symmetry is broken concomitantly with the FE phase as speculated
before. According to the models suggested by others \cite{9,10},
displacements of Mn$^{3+}$ are in \textit{ab}-plane. While
\textit{b}-axis components of the atomic displacements mainly
contribute to  $\emph{P}$, \textit{a}-axis components enable the
emergence of \textit{I}$_{(300)}$. If the atomic displacements
correspond to the periodic lattice modulations, it is expected that
both $\emph{P}^2$ and \textit{I}$_{(300)}$ are proportional to
\textit{I}$_c$, as shown in Fig. 2 (b). (The spontaneous
polarization data are taken from Ref. 4 and are shifted in order to
get the same values for \textit{T}$_{c1}$.) It confirms that
spontaneous polarization is due to the atomic displacements
 driven by magnetic orders: a direct
crystallographic evidence of exchange striction  as the origin of
ferroelectricity in the material \cite{8,9,10,12}. Also it is noted
that  \textit{I}$_{(300)}$ drops abruptly at \textit{T}$_{c2}$ and
has a broad minimum around \textit{T}$_{c3}$.

\begin{figure}[b]
\begin{center}
\includegraphics [width=8.5cm] {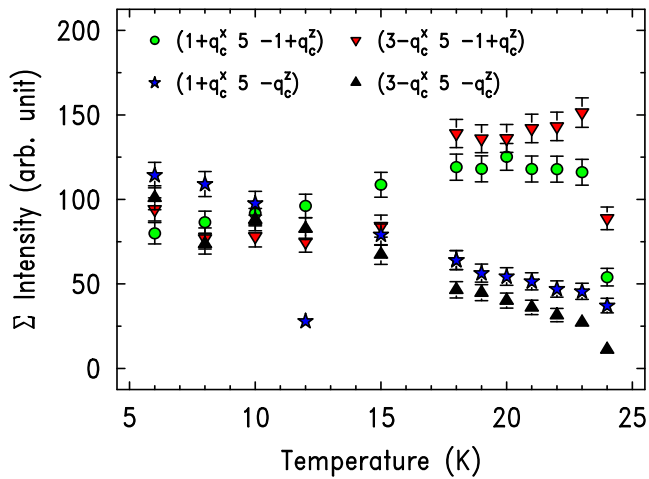}
\caption {(Color online) Temperature dependences of the
ICM lattice modulation peak intensities.}
\end{center}
\end{figure}

\begin{figure}[b]
\begin{center}
\includegraphics [width=8.5cm] {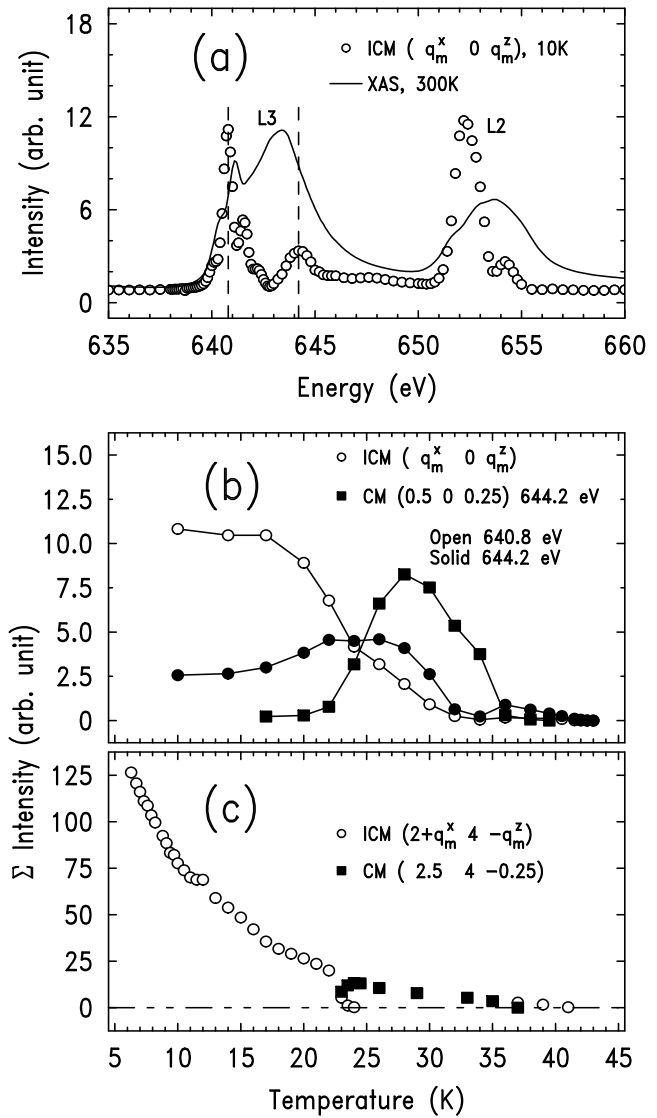}
\caption {(a) Energy profiles of the ICM magnetic peaks (circle)and
XAS (solid line) around
Mn \textit{L$_{2,3}$}-edges. Vertical broken lines correspond to
640.8 eV and 644.2 eV, respectively. (b) Temperature dependences of the
ICM (circle) and the CM (square) magnetic peaks. Open (Solid) symbols denote the data taken E
= 640.8 eV (644.2 eV), respectively. (c) Temperature dependences of the
ICM (open circle) and the CM (solid square) magnetic peak at Tb \textit{L}$_3$-edge.}
\end{center}
\end{figure}

Though many interesting ME phenomena have been reported in the
LT-ICM phases  below \textit{T}$_{c2}$ \cite{4,11,18,19}, their
basic mechanisms still remain to be understood. Since the lattice
modulations reflect basic ME natures,
 temperature dependences
below \textit{T}$_{c2}$ of integrated intensities were measured at
the four split ICM peak positions illustrated in Fig. 1 (a). From the
results displayed in Fig. 3, it is clear that
temperature dependences of all four peaks cannot be described by a
single order parameter, implying  the presence of various magnetic
orders having the same \textit{\textbf{q}$_m$}'s but different
temperature dependences.

To investigate different magnetic orders, we performed resonant
x-ray magnetic scattering measurements at Mn \textit{L}-, Tb
\textit{L$_3$}- and \textit{M$_5$}-edges. Figure 4 (a) shows energy
profiles around Mn \textit{L}-edge of magnetic satellites at 10 K
and x-ray absorption spectroscopy (XAS) at room temperature.
Magnetic peaks and XAS data clearly show resonances at both Mn
\textit{L$_{2}$}- and \textit{L$_{3}$}-edges. XAS results show broad
peaks containing contributions from the multiplet states of 3$d$
electrons of Mn$^{3+}$  and Mn$^{4+}$  ions. Magnetic satellites
show relatively sharp double peaks at both Mn \textit{L}-edges. The
sharp resonances represent different multiplet states of Mn 3$d$
electrons  including charge transfer excitations, while Mn ions are
expected to be in the high-spin configurations with all the 3$d$
electron spins aligned  parallel. Therefore, although the resonances
do not have one-to-one correspondences with the magnetic orders of
Mn ions, changes in the resonances at magnetic satellites reflect
the changes in spin ordering which are periodically modulated with
the wave vector \textit{\textbf{q}$_m$}. Temperature dependences of
x-ray intensities at the ICM peak of \textit{\textbf{Q}$_m$} =
(\textit{q$_m^x$} 0 \textit{q$_m^z$}) were measured at the two
resonances, 640.8 and 644.2 eV. The results are presented in Fig. 4
(b). Data for a CM peak of \textit{\textbf{Q}$_m$} = (0.5 0 0.25) at
the resonance of 644.2 eV are presented together.  It is clear that,
above 15 K, intensities of each resonance  have different
temperature dependences from each other. Though the origin of the
anomalous temperature dependences is not understood in detail, it
reflects complicated natures of magnetic moments of Mn ions under
the frustrated configuration.

Magnetic ordering of Tb$^{3+}$ ions was investigated with resonant
x-ray scattering measurements at Tb \textit{L$_3$}-edge. Figure 4
(c) shows that ordering temperature of Tb magnetic moments is the
same with that of Mn, \textit{T}$_N$, which is consistent with
neutron scattering results \cite{9}. The modulation wave vector of
Tb magnetic order is the same with the values of
\textit{\textbf{q}$_m$} measured in nonresonant x-ray scattering.
Soft x-ray magnetic scattering measurements were also performed at
Tb \textit{M}$_5$-edge and the result not shown here confirms that
observed x-ray intensities in Fig. 4 (c) reflect magnetic order of
Tb $4f$ electrons which grows monotonically below \textit{T}$_N$.

From the results shown in Fig. 4 (b) and (c), it is clear that there
exist multiple magnetic order parameters having the same
\textit{\textbf{q}$_m$}'s but different temperature dependences. The
contributing portions of each magnetic order to scattering factors
of magnetic satellites  are different depending on
\textit{\textbf{Q}$_m$}(= \textit{\textbf{Q}$_{Bragg}$} +
\textit{\textbf{q}$_m$}), and it results in different temperature
dependence for each magnetic peak and its corresponding lattice
modulation peak intensities, which explains the temperature
dependences presented in Fig. 3.

Since the magnetic orders are located under the spin frustrated geometry,
it is reasonable to suppose that phase-slips take place  due to
competitions between the magnetic orders, as their order parameters grow
with different temperature dependences.
 The discommensuration
results in the transition to the LT-ICM phase. Anti-phase domain
walls for the phase slips are consistent with the aforementioned
long-period CM modulations below \textit{T}$_{c3}$. Assuming the
model suggested by others \cite{9}, atomic displacements are canted
antiferroelectric type. Across an anti-phase domain wall, directions
of the atomic displacements and the spontaneous polarizations are
reversed. Therefore, not only the polarizations from domains
separated by the domain wall cancel each other but also x-ray
scattering amplitudes for the (3 0 0) Bragg peak are canceled due to
the crystal symmetry. Then, only remnants resulting from unequal
populations of the domains contribute to $\emph{P}$ and
\textit{I$_{(300)}$}. Since a density of the domain walls determines
\textit{\textbf{q}$_m$}, temperature dependences of $\emph{P}$,
\textit{I$_{(300)}$} and \textit{\textbf{q}$_m$} down to
\textit{T}$_{c3}$ can be explained consistently in terms of CM
modulations with the anti-phase domain walls. This indicates that CM
modulations are preferred as its low temperature ground state. Then,
the low temperature phase seems to have a higher entropy due to the
domain walls than the high temperature CM phase, violating the
entropy rule. However, due to the geometrical frustration and the
presence of multiple magnetic orders many different energy scales
can exist. The complicated temperature dependences of the magnetic
orders in Fig. 4(c) reflect the presence of the different energy
scales. Smaller energy scales become important at low temperatures
and induce discommensuration. Upturns of the electrical polarization
and \textit{I$_{(300)}$}
 below \textit{T}$_{c3}$ are attributed to lattice modulations
enhanced by increasing Tb magnetic moments, which is consistent with
results of others demonstrating couplings between Tb moments and
lattices \cite{18,19,20}.


In summary, we have shown that exchange striction is the driving
mechanism for the magnetoelectricity in the material. The same
temperature dependences of x-ray intensities at a (3 0 0) forbidden
Bragg peak and a lattice modulation peak in the CM FE phase,
together with observation of the relation, \textit{\textbf{q}$_c$} =
2\textit{\textbf{q}$_m$}, demonstrate that spontaneous electric
polarization is due to atomic displacements driven by the exchange
striction of magnetic orders. Resonant x-ray magnetic scattering
results confirm the presence of multiple magnetic orders having
different temperature dependences. The CM to LT-ICM phase transition
is attributed to discommensuration through phase slipping in the
competing magnetic orders
 in the frustrated configurations.
Temperature dependences of \textit{\textbf{q}$_m$}, $\emph{P}$ and
\textit{I$_{(300)}$} in the LT-ICM phase are explained in terms of
the CM modulations with anti-phase domain walls.

We thank D.J. Huang for the useful discussions.
This work was supported by the KOSEF through the eSSC at POSTECH,
and by MOHRE through BK-21 program.
The experiments at the PLS were supported by the POSTECH
Foundation and MOST.
Use of the Advanced Photon Source (APS) was supported by the U.S. Department of Energy,
Office of Science, Office of Basic Energy Sciences, under Contract No.
W-31-109-Eng-38.
The Midwest Universities Collaborative Access Team (MUCAT) sector at the APS is
supported by the U.S. Department of Energy,
Office of Science, Office of Basic Energy Sciences, through the Ames Laboratory
under contract No. DE-AC02-07CH11358.
Work at Rutgers was supported by NSF-DMR-0520471.


\end{document}